\documentclass{aa}  
\usepackage{lipsum} 

\usepackage{graphicx}
\usepackage{txfonts}
\newcommand{\Msun}{\mbox{${M}_{\odot}$}}
 \newcommand{\Zsun}{\mbox{${Z}_{\odot}$}}
\newcommand{\Mi}{\mbox{$M_{\rm i}$}}

\newcommand{\Zcr}{\mbox{$Z_{\rm cr}$}}

\newcommand{\Mcot}{\mbox{$M5{\rm \omega}3$}}
\newcommand{\Mcoq}{\mbox{$M5{\rm \omega}4$}}
\newcommand{\Msot}{\mbox{$M6{\rm \omega}3$}}
\newcommand{\Msoq}{\mbox{$M6{\rm \omega}4$}}

\newcommand{\MUP}{\mbox{${M_{\rm UP}}$}}
\newcommand{\xUP}{\mbox{${x_{\rm UP}}$}}

\newcommand{\Mchar}{\mbox{${M}_{\rm char}$}}

\usepackage[normalem]{ulem}
\begin{document}

\title{On the Contribution of Very Massive Stars to the Sulfur Abundance in Star-Forming Galaxies: the Role of PISN}

\titlerunning{Contribution of Very Massive Stars to the Sulfur Abundance in Star-Forming Galaxies}
\authorrunning{S. Goswami et al.}

    \author{S. Goswami
          \inst{1}
          \and
        J. M. Vilchez\inst{2} 
        \and
        B. Pérez-Díaz\inst{2}
        \and
        L. Silva
           \inst{3,4}
           \and
           A. Bressan
           \inst{5}
           \and 
           E. Pérez-Montero
           \inst{2}}
        
    \institute{
        Department of Applied Physics, University of Cádiz, Campus of
Puerto Real, 11510 Cádiz, Spain\\
\email{sabyasachi.goswami@uca.es  
        }
        \and
    Instituto de Astrofísica de Andalucía - CSIC, Apdo 3004, E-18080, Granada, Spain
    \and
        INAF-OATs, Via G. B. Tiepolo 11, I-34143 Trieste, Italy
        \and
        IFPU - Institute for Fundamental Physics of the Universe, Via Beirut 2, 34014 Trieste, Italy
        \and
         SISSA, Via Bonomea 265, I-34136 Trieste, Italy\\
   }
   %

 
  \abstract
   {Recent work presented increasing evidence of high, non-constant S/O abundance ratios observed in star-forming metal-poor galaxies, showing deviations from the constant canonical S/O across a large range of O/H abundance. Similar peculiar high Fe/O ratios have been also recently detected.
   }
    {We investigate whether these high S/O ratios at low metallicities could be explained taking into consideration the process of Pair Instability Supernovae (PISN) in chemical modelling through which similar behaviour observed for Fe/O ratios was successfully reproduced.
    }
  {We use chemical evolution models which take into account the stages of PISN  in the yields published by \cite{Goswami2022}, and adopt a suitable initial mass function (IMF) to characterize this evolutionary stage appropriately. }
  {The peculiar high values and the behaviour of the observed S/O versus O/H relation  can be reproduced when the ejecta of very massive stars that go through the process of  PISN are taken into account. Additionally, a bi-modal top-heavy IMF and an initial strong burst of star formation are required to attain the reported high S/O values}
{We show that the role of very massive stars going through the process of PISN should be taken into account when explaining the chemical enrichment of sulfur and oxygen in metal-poor star-forming regions.}
\keywords{galaxies: abundances – galaxies: starburst – galaxies: formation – stars: massive – galaxies: structure}
   \maketitle

%

\section{Introduction}
\label{intro}
Sulfur (S) is produced in massive stars where helium is fused into heavier elements, including S, during the final stages of the star's life, before the supernova explosion. An oxygen (O) convective core is formed during the hydrostatic burning of neon, which leads to the formation of $\alpha$-elements till S. Sulfur is additionally produced during explosive O burning during the supernova (Type II) explosion. Since oxygen is produced through the same processes
 the study of the S/O ratio can give valuable hints on the chemical evolution, star formation history and Initial Mass Function (IMF). 
There is a collection of studies where sulfur and oxygen abundances have been derived and studied in the context of star-forming galaxies and giant H~II regions (e.g. \citealt{Pagel1978,Vilchez1988,Garnett1997,Perez2006,Lopez2009,Berg2013,Dors2016,Fernandez2019,Arellano2020,Dors2023}). However, the debate continues today regarding the S/O relationship with metallicity as usually measured in terms of (O/H). Several studies \citep{Garnett1997,Izotov2006,Guseva2011,Berg2020,Rogers2021} have shown that the S/O ratio remains constant with metallicity traced by O/H. Since sulfur and oxygen are produced through the same processes described above, their ratio is expected to be constant as long as they are produced by similar mass ranges of stars. On the contrary, some studies have pointed out that there could be a variation of S/O with decreasing metallicity (e.g. \citealt{Vilchez1988,Dors2016,Diaz2022}). Recently, \cite{Diaz2022} measured sulfur and oxygen abundances in a large sample of galaxies dominated by young, massive stars, finding a clear bimodality in the trend of the S/O ratio with metallicity. In their sample, HII regions in compact dwarf galaxies (HII Gal), that are dominated by
strong starbursts show S/O ratios mostly below or around solar S/O,  and an increasing trend of the S/O ratio with increasing metallicity (O/H). On the contrary, Giant diffuse HII regions (DHR),  mostly in irregular and spiral galaxies,  exhibit
the opposite trend with high S/O ratios at low metallicities and a decreasing S/O
ratio at increasing O/H.
The most peculiar aspect is the extremely high S/O
ratio observed in DHRs at low metallicities( 12
+ log(O/H) ~ 7.0), for which a theoretical explanation is still lacking.

Similarly, high Fe/O ratios at low 12+log(O/H) values have been recently reported in extremely metal-poor star-forming dwarf galaxies (with 12+log(O/H)$\leq$7.69; \citealt{Kojima2020}). The high Fe/O ratios observed in  these metal-poor galaxies were explained in the work by
\cite{Goswami2022} by considering in their chemical evolution models very massive stars that undergo processes such as
PPISN/PISN (Pulsational PISN) through their later evolutionary stages. 
Sulfur is produced through pair instability supernova which occurs in very massive stars, with initial masses greater than around 130 M$\sun$. The PISN stage strongly depends on the mass of the He core at the end of the H burning phase \citep{Heger_Woosley2002y,Takahashi2018ApJ857}, which produces large amounts of elements such as Fe, S, O, among others. \cite{Takahashi2018ApJ857} in their work obtained new yields from PISN nucleosynthesis for rotating and non-rotating zero-metallicity stars. They compared their theoretical models with the abundances of metal-poor stars from the SAGA database, and they did not find evidence of PISN metal-poor stars in the sample based on [Na/Mg] and [Ca/Mg] abundances. Since then, there have been multiple observations showing high Fe/O ratios \citep{Kojima2020}, and low [Mg/Fe] ratios \citep{Yoshii2022}  both of which have been linked to evidence of PISN nucleosynthesis enrichment. Additionally, \cite{2023NaturX} reported abundances in very metal-poor(VMP) stars with extremely low sodium and cobalt abundances, which provide a chemical signature of PISN. One of the major difficulties in getting a direct signature is that PISN explosions do not leave behind any remnants like CCSN does, and hence the tracers of such explosions can only be indirectly found.  Chemical abundances of various elements, if available, can provide us with crucial information that may allow us to differentiate between multiple possible nucleosynthesis channels, such as Hypernovae (HNe) depending on which elements are produced.  PISN yields have been used as a possible channel in previous studies to analyse the chemical evolution of different galaxies \citep{Isobe_2022,Watan2023}. It has been shown that this stage is expected to
be limited up to a  threshold metallicity  of Z$_{cr}\lesssim 10^{-4} Z_{\odot}$  or  Z$_{cr}\lesssim 10^{-2} Z_{\odot}$
depending on efficient or inefficient dust cooling, respectively q\citep{Schneider2006MNRAS,salvadorieial2008MNRAS}. However, with current models, the PISN stage is suggested to be important till Z$\sim$ 0.5 \Zsun\ \citep{Kozyreva2014a, Langer2012, Costaetal2020}.

Observationally, the VLT Tarantula Survey \citep{Schneider2006MNRAS}  which studies the 30 Doradus region in the Large Magellanic Cloud (LMC) inferred very young ages (1-6 Myr)  stars with masses of up to 200 M$_{\rm \odot}$.  Additionally, \cite{Crowtheretal2016MNRAS} estimated star masses that could be as high as 300 M$_{\rm \odot}$ in the same region. Moreover, the IMF derived in specific regions of NGC~2070 shows an upper mass slope of $x = 0.65$ and an overall slope of $x = 0.9$. 
In order to investigate the high S/O ratio observed in some metal-poor star-forming galaxies, we have computed new chemical evolution models for these objects, including
the PISN yields.
A comparison of these models with observations can thus provide valuable  information,
firstly, concerning the nature of the chemical enrichment process at very low metallicity, which is especially interesting for the evolution of galaxies in the early universe,
and, secondly, on the nature of the IMF  and its possible deviations
from universality.

In summary, sulfur is a very relevant element to be studied in the regions hosting very massive star formation, especially significant for the studies of galaxies in the early universe. Its unique nucleosynthesis history and the new data coming from the JWST for low metallicity galaxies can allow us to understand the viable role played by very massive stars in the chemical evolution of galaxies, and, eventually, help us decipher the properties of the first stars. Thus, the primary goal of this work is to study whether accounting for the PISN stage of very massive stars in chemical evolution modelling could be a possible channel to replicate the unusually high S/O ratios observed, as shown by the DHR sample of \cite{Diaz2022} at very low metallicity (12+ log (O/H) $\lessapprox$ 7.5). As previously mentioned, PISN is primarily supposed to occur in low-metallicity environments, where high S/O ratios have been observed. Combined with the fact that PISNs produce a large amount of S, this study of sulphur abundance at low metallicities can offer us a  unique tracer to reconstruct the early universe's nucleosynthesis history and to comprehend the characteristics of its very massive stars. \\

This paper is structured as follows. In Section \ref{method} we describe the chemical evolution model used in this work. In particular, we describe the PISN yields of oxygen, sulfur and iron.  In Section \ref{observational}, we present the sample of observational data analysed in this work. 
In Section \ref{evolu}, we compare the observations to the predictions of the most suitable chemical evolution models stressing the derived evolutionary constraints. 
Finally, in Section \ref{conclu}, we summarise our conclusions and outlook.

\section{Methodology}
\label{method}
In this section, we describe the chemical evolution model we have developed to be used as a tool to understand the high S/O ratios at low metallicity, as shown by recent studies. We use the 
code \texttt{CHE-EVO} \citep{Silva1998} which computes  one-zone open chemical evolution models considering the
time evolution of the gas elemental abundances, and includes the
infall of primordial gas.
This code has been used in several contexts to provide the input star formation and metallicity histories to interpret the spectrophotometric evolution of both normal and starburst galaxies 
\citep[e.g.][]{Vega2008,Fontanot2009,Silva2011,Lofaro2013,Lofaro2015,Hunt2019}. 
The basic equation used in this code can be written as follows:
\begin{equation}
\dot{M}_{{\rm g},j}=    \dot{M}_{{\rm g},j}^{\rm Inf} - \dot{M}_{{\rm g},j}^{\rm SF} +  \dot{M}_{{\rm g},j}^{\rm FB} 
\label{eq_chem_ev}
\end{equation}
where, for the element \textit{j}, the first term on the right, $\dot{M}_{{\rm g},j}^{\rm Inf}$ corresponds to the infall rate of pristine material.  $\dot{M}_{{\rm g},j}^{\rm SF}$ ,  represents the rate of gas consumption by star formation,  and $\dot{M}_{{\rm g},j}^{\rm FB}$ refers to the rate of gas return to the interstellar medium (ISM) by dying stars.
The latter term also includes the contribution of type Ia supernovae (SNIa), whose rate is adjusted with the parameter
${A_{\rm SNIa}}$ which corresponds to the fraction of binaries with system masses between  3 M$_{\rm \odot}$ and 16 M$_{\rm \odot}$ and the right properties to give rise to SNIa \citep{Matteucci1986}.
\\We used the Schmidt-Kennicutt law \citep{Kenni1998} to model the star formation rate (SFR):
\begin{equation}
\psi(t)= \nu \,  M_{\rm g}(t)^{k} 
\label{eq_SF_law}
\end{equation}
where $\nu$  is the efficiency of star formation, $M_{\rm g}$ is the mass of the gas, and $k$ is the exponent of the star formation law. The gas infall law is assumed to be exponential \citep[e.g.][]{Grisoni_2017,Grisoni2018}, with an e-folding timescale of $\tau_{\rm inf}$. The IMF can be written as follows:
\begin{equation}
\phi(M_{\rm{i}}) =  \frac{dn}{d\log(M_{\rm{i}})} \propto M_{\rm{i}}^{-x}
\label{imfkroupa}
\end{equation}

\subsection{Stellar Yields}
\label{yields}
We use in this work the stellar yield compilation as in \cite{Goswami2022}. Here we give a brief overview and the reader is addressed to the previous work for a more thorough review. 
For low- and intermediate-mass stars ($M_i <$ 8 $M_{\rm \odot}$), single stars and binary systems that give rise to SNe Ia are characterized by different yields \citep{Matteucci1986}.
\\
For single stars with initial masses $M_i <$ 6 $M_{\rm \odot}$ we adopt the yields from calibrated Asymptotic Giant Branch (AGB) models by \cite{Marigo2020NatAs}. For $6\Msun < M_i < 8 \Msun $ we use super-AGB yields from \cite{Ritter_etal18}. 
For SNIa 
we consider the yields provided by \cite{1999ApJS..125..439I}.
For $8$ $M_{\rm \odot} \leq M_i \leq$ $350$ $M_{\rm \odot}$ we adopt the yield compilation by \cite{Goswami2022}  for massive stars and very massive objects (VMO). \\

\begin{figure} 
\centering
\resizebox{1.0\hsize}{!}{\includegraphics[angle=0]
{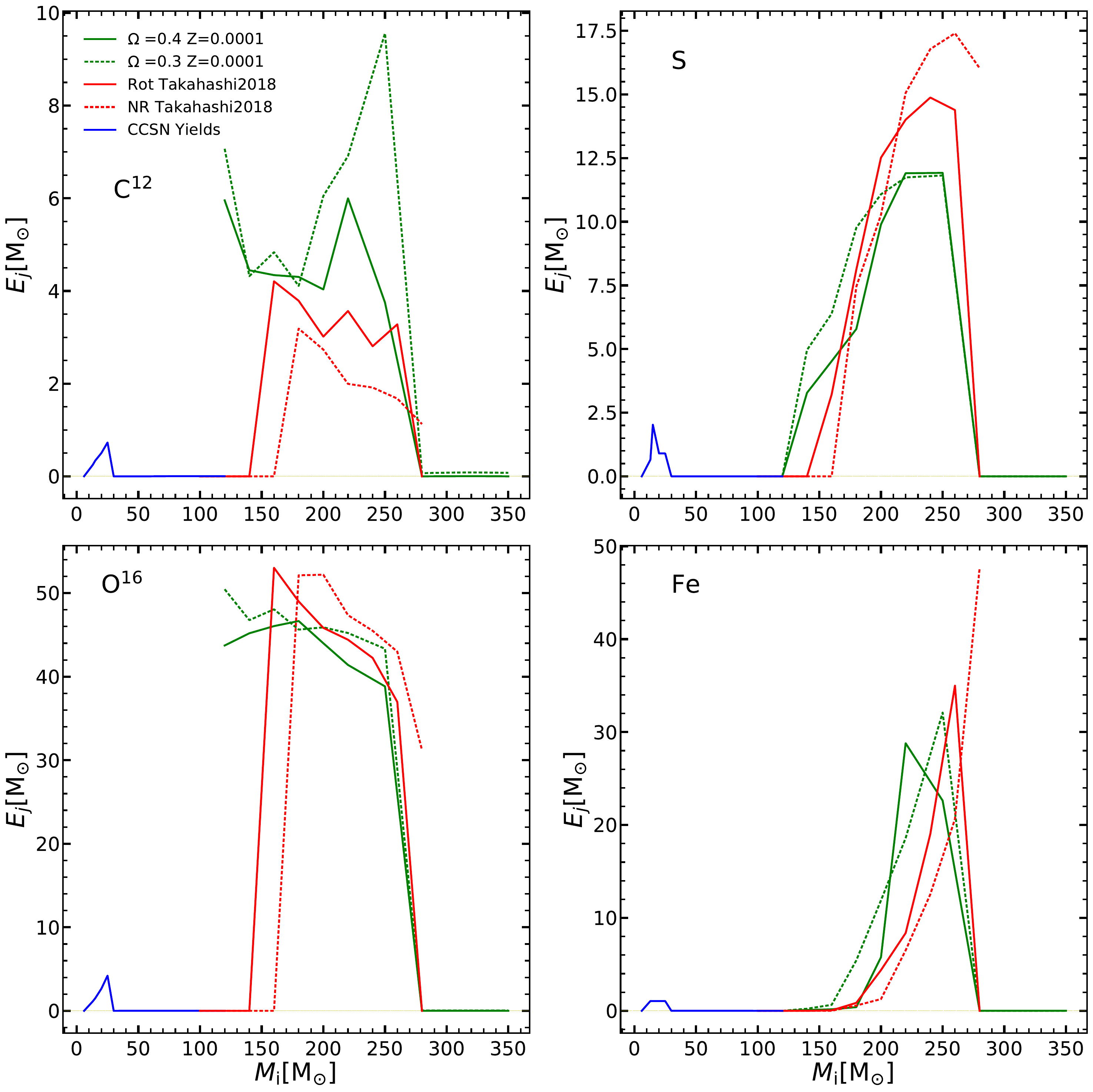}}  

\caption{The relation between newly produced ejecta ($E_{j}$) in the $y$-axis  of the elements C$^{12}$, O$^{16}$, S  and Fe versus the initial mass  ($M_{\mathrm{i}}$) in the $x$-axis  from Z=0.0001 rotational models of \cite{Goswami2022} in green (solid, and dotted lines for 2 rotation rate values), based on \cite{Heger_Woosley2002y} pure He non-rotating models.CCSN yields are shown in blue.  In red,  yields of rotating (solid) and non-rotating (dotted) models by \cite{Takahashi2018ApJ857} for zero metallicity stars are shown.  Since the actual S and Fe CCSN yields are small, they are shown multiplied by a factor of 15 for the sake of clarity. }
\label{pisnyields} 
\end{figure}

A comparison of PISN ejecta with other models in the literature has been shown in Figure~\ref{pisnyields}. PISN model predictions from \cite{Goswami2022} have been shown in green, which includes rotation. These ejecta have been computed by matching the  \cite{Heger_Woosley2002y} models, which do not include rotation originally with the corresponding PARSEC \citep{Bressan2012,Costaetal2020} rotating models.  Rotating and non-rotating models from \cite{Takahashi2018ApJ857} are shown in red solid and dotted, respectively. Finally, in magenta, the rotation included CCSN yields derived by \cite{Limongi_etal18} and used in \cite{Goswami2022} are shown. This allows us to explicitly display the difference in the production of the relevant elements through the two channels, i.e., CCSN and PISN at low metallicities. Looking at the PISN yields in green and red, we notice that the production of sulfur increases as the initial mass increases. As discussed in Sect. \ref{intro} the PISN yields strongly depend on the mass of the helium core, nevertheless for the sake of simplicity, here we show the yields against the initial masses. This also allows us to compare the PISN yields with the CCSN yields simultaneously. The trend of increased production is also true for iron, where stars with more massive initial masses (as a result of a more massive helium core) produce more iron. However, carbon and oxygen have the opposite trend, as the ejecta of both carbon and oxygen starts to decrease as the mass of the helium core increases. The effect of rotation is also more clearly seen in the case of carbon and oxygen where the production of these elements grows with higher values of rotational velocities.  Finally, we can see that rotation does not play an important role in sulfur production. Comparing the PISN yields to the CCSN yields for the same metallicity $Z_{i}=0.0001$ shown in blue, we can notice that there is relatively minimal sulfur and iron production through the CCSN channel at low metallicities. The production of carbon and oxygen is higher through the CCSN as compared to S and Fe; however, the relative difference between the CCSN and PISN yields for all four elements is significant in the low metallicities regime. 

\begin{figure} 
\centering
\resizebox{0.6\hsize}{!}{\includegraphics[angle=0]{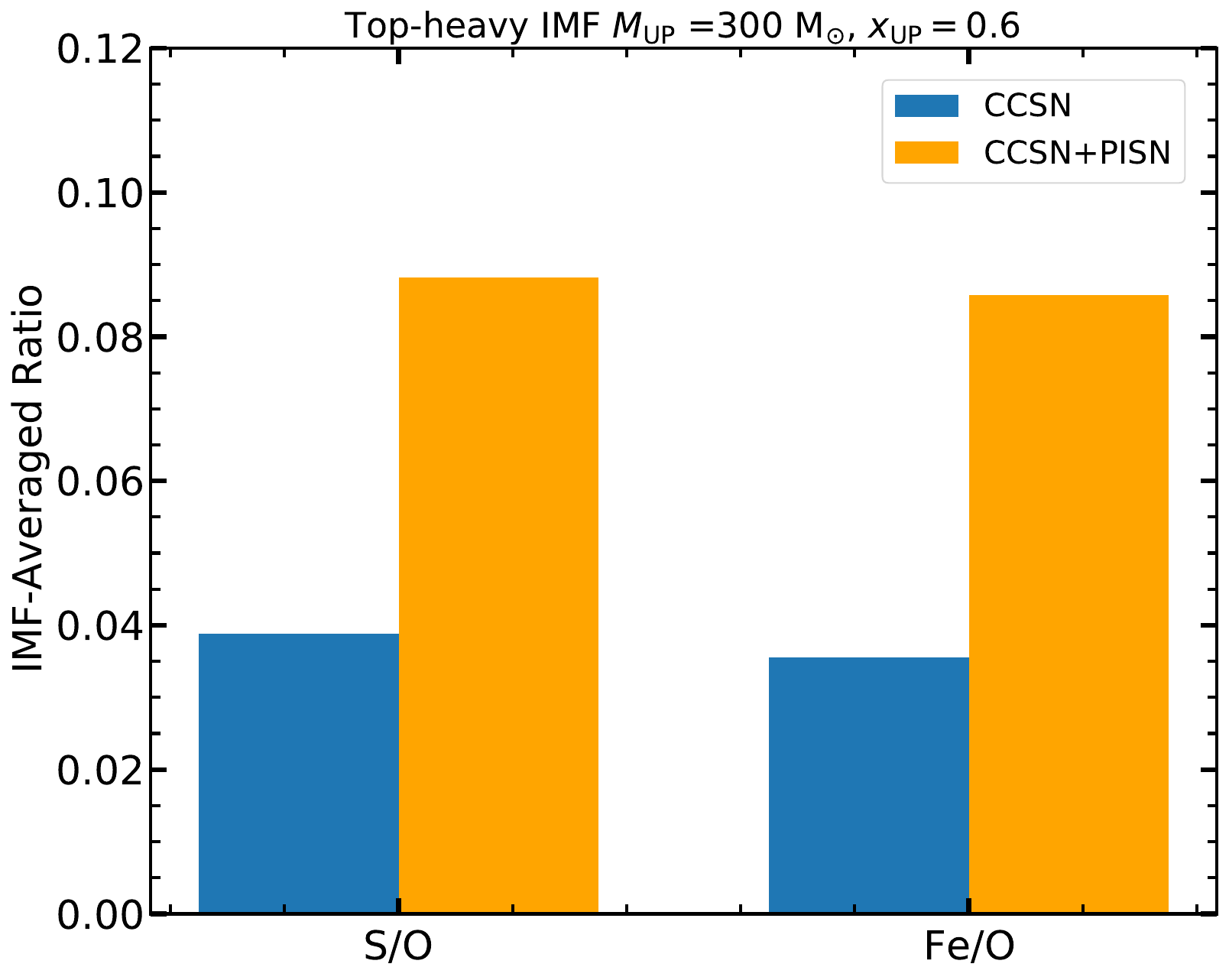}}  
\resizebox{0.6\hsize}{!}{\includegraphics[angle=0]{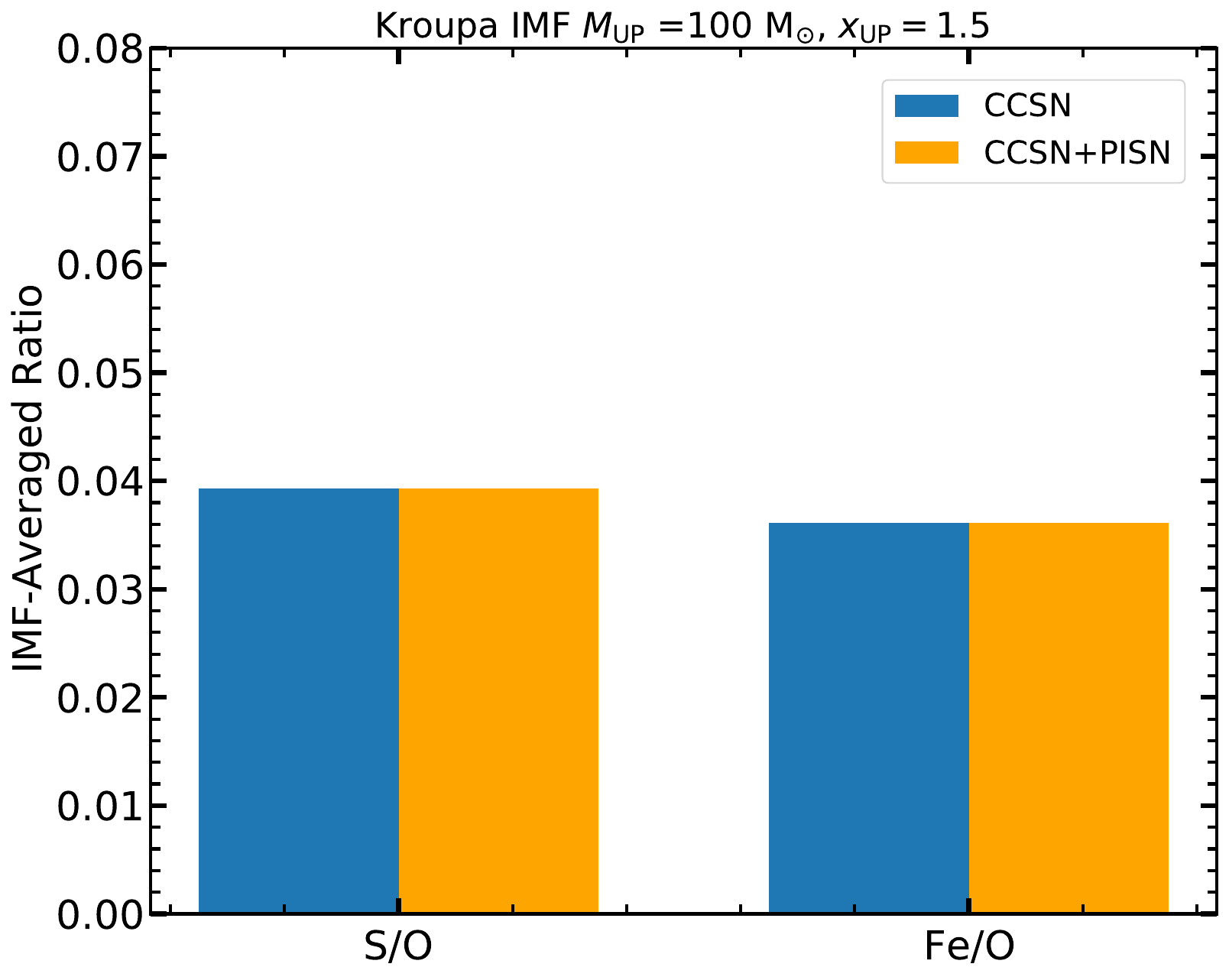}}
\caption{IMF-averaged ratios of S/O and Fe/O shown for two IMFs: a top-heavy IMF with a 300 \Msun\ upper  mass limit (PISN stage is accounted for) and a canonical Kroupa IMF with a 100 \Msun\  limit for the yields sets adopted in this work.} 
\label{imfyields} 
\end{figure}

To illustrate this better, in Fig.~\ref{imfyields} we show the IMF-averaged S/O and Fe/O ratios for the CCSN and PISN yields used in this work. We present the element ratios for two distinct IMFs to demonstrate how, when relying solely on CCSN yields and a canonical IMF, the S/O and Fe/O ratios obtained are not as high as they would be if PISN were also taken into consideration. The top-panel of Fig. \ref{imfyields} shows the element ratios averaging over the IMF  obtained from the CCSN and PISN yields using a top-heavy IMF that has an upper mass limit of 300 \Msun\ and a slope of \xUP\ =0.6. Since it has a high upper mass limit and the stepper slope increases the number of such very massive stars.The PISN stage, due to explosive O burning, produces
higher amounts of S, Fe, that are taken into account, which causes the
S/O and Fe/O ratios to be high. O production is lower in the high-mass PISN regime as compared to low-mass PISN since O is fused to other heavier elements. In the bottom panel, the element ratios averaged over a canonical Kroupa IMF with an upper mass limit of \MUP= 100 \Msun\ have been shown. Because of the upper mass cut, the very massive stars going through the PISN stage are not accounted for, so the contribution of these ratios comes primarily from CCSN yields, which is why the two bars match each other. 


\begin{figure*} 
\centering
\resizebox{0.41\hsize}{!}{\includegraphics[angle=0]{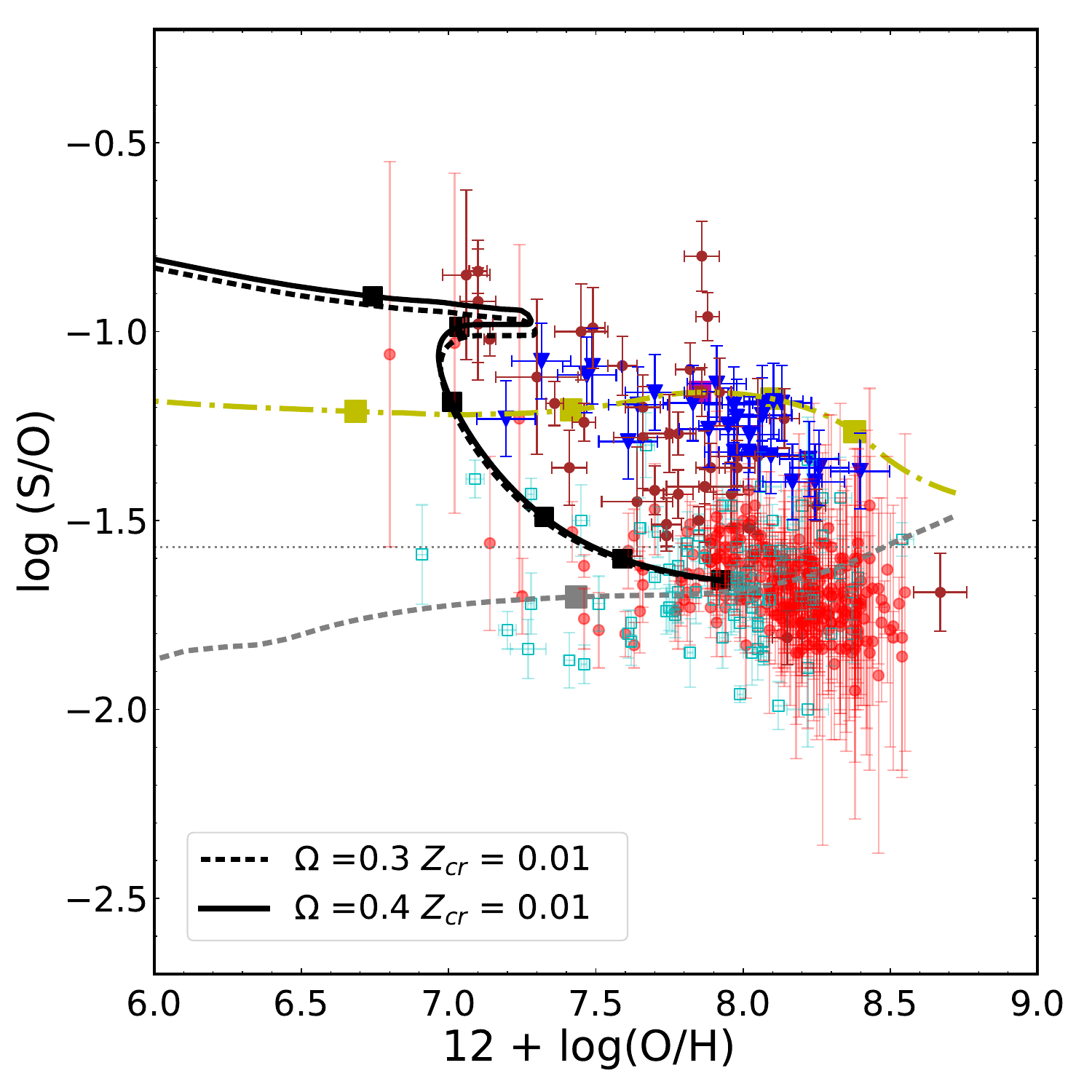}}  
\resizebox{0.41\hsize}{!}{\includegraphics[angle=0]{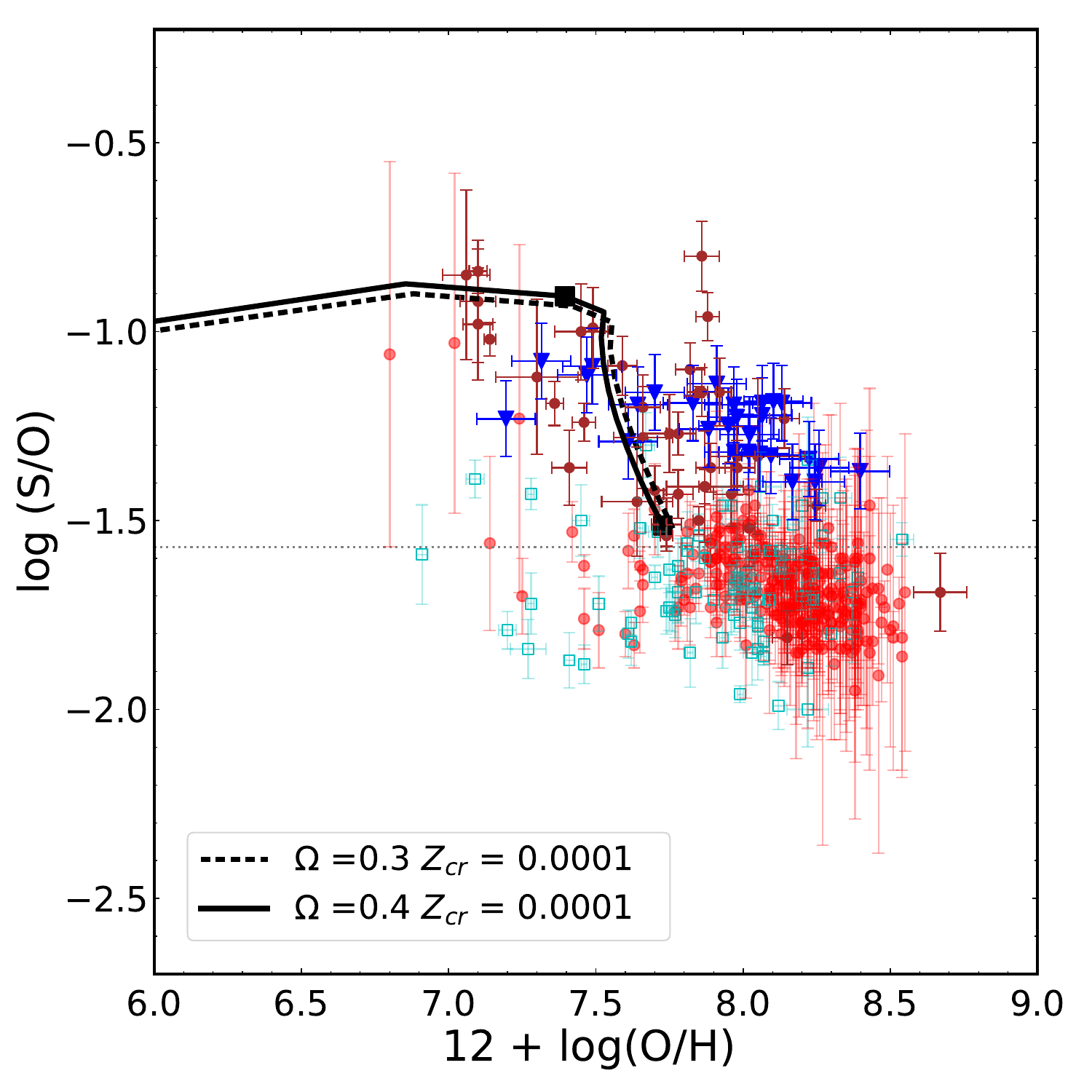}}
\caption{S/O ratio for two different models with PISN: M5 (\Mcot\ and \Mcoq)  in the left panel and M6 (\Msot\ and \Msoq) in the right panel as a function of the oxygen abundance 12+log(O/H). The black dotted and solid lines denote the two different values of rotational parameters used. The red points are from \cite{Izotov2006} and the blue triangles are from \cite{Yates2020}. The brown and cyan points correspond to the DHR and HII Gal subsamples of   \cite{Diaz2022} respectively. Model M1, in grey dashed line, is a chemical evolution model representative of the MW disc \citep{Goswami2020} where the grey square corresponds to an age of 200Myr, and the M3 model predictions are shown in yellow dashed-dotted line. The six squares on models M5 and M3 represent the model's age at 5, 15, 30, 60, 100, and 200 Myr. For M6, only ages corresponding to 5 and 15 Myr are shown by the two squares. The values adopted for critical metallicity \Zcr\ and rotational parameter $\Omega$ are quoted for the two models. The grey dot-dashed line denotes the solar S/O ratio. See the text and Table \ref{parameter} for details.} 
\label{soratio} 
\end{figure*}

\section{Observational data}
\label{observational}
The abundance data sample used in this work has been taken from the extensive work by \cite{Izotov2006} and \cite{Diaz2022}, with a selection also of H II regions from \cite{Yates2020}. \cite{Izotov2006} examined the chemical composition of the emission line galaxies of the Sloan Digital Sky Survey (SDSS) Data Release 3. They measured the [SII]$\lambda$$\lambda$6717,31 and [SIII]$\lambda$6312 lines\footnote{From S$^{+}$ and S$^{++}$ ions, typically representing common and abundant ionic species of sulfur in star-forming regions} of sulfur, and used the direct method for the derivation of sulfur and oxygen abundances. The sulfur abundances they derived in most of these galaxies exhibit a similar pattern to that obtained for nearby galaxies in the high metallicity regime; however, at the low metallicity end (12+log(O/H) $\lessapprox$ 7.2) the S/O ratios derived 
exceed the canonical solar value (log(S/O)=-1.57; \citealp[]{Asplund2009}).  \cite{Diaz2022} in their innovative work derived sulfur abundances for a large sample of HII regions in irregular and spiral galaxies including the Magellanic Clouds (Giant Diffuse HII Regions; DHR subsample) and dwarf galaxies with a prominent starburst (HII Gal subsample), using measurements of [SII]$\lambda$$\lambda$6717,31 and [SIII]$\lambda$$\lambda$9069,9532 lines from the literature and re-calculating the abundances following the direct method, based on the derivation of the electron temperature.  Similarly,  abundances derived with the direct method for a set of giant H II regions are also included from \cite{Yates2020} who measured also the [SII] and [SIII] lines. Peculiarly high ratios of S/O in the lowest metallicity range were found by \cite{Diaz2022} in their DHR sample, and confirmed using the \cite{Yates2020} data. Both these samples follow a similar trend of high S/O ratios at low O/H and then a decrease in S/O with the increase in metallicity with some dispersion at high O/H. However, in the HII Gal sample of \cite{Diaz2022} and the majority of the \cite{Izotov2006} sample the S/O ratios are around the solar S/O value. In this work, we concentrate on the more anomalous high S/O ratios at low metallicities seen in the DHR sample of \cite{Diaz2022}, for which there is currently no theoretical explanation. 
It is important to emphasize here that invoking possible depletion of oxygen in dust grains ($\approx$0.08 dex for this metallicity; \citealp[]{PeimbertPeimbert2010}) could not explain the observed trend, especially at such low abundance. \\
We note that the three abundance studies selected above, irrespective of their  specific abundance derivation techniques, reported high S/O values independently, which our models reproduce for a number of star-forming
objects at low metallicity. Performing a study for a larger complete sample with fully homogeneously derived 
sulfur abundances is out of the scope of this work.

In Figure~\ref{soratio} we show the log(S/O) ratio versus oxygen abundance, 12+log(O/H). The red data points correspond to the sample from \cite{Izotov2006} while the brown and cyan points represent the abundances derived from \cite{Diaz2022} for the DHR and HII Gal samples respectively, and the blue triangles correspond to star-forming regions from  \cite{Yates2020}. As it can be seen in Figure~\ref{soratio}, as the O/H abundance decreases, the S/O ratio starts to increase in the DHR and the \cite{Yates2020} sample shown in the brown and blue points respectively and clearly deviates from the constant solar ratio suggested by various studies mentioned in Sect.~\ref{intro}. On the other hand, the data is more dispersed around the solar S/O ratio for the HII Gal and most of the \cite{Izotov2006} sample, which are indicated by cyan and red dots, respectively. In the following section, we explore whether chemical evolution models that account for the PISN stage can replicate the high S/O ratios at low metallicity demonstrated by the DHR sample.
\section{Results}
\label{evolu}
 We have used chemical evolution models to study the evolutionary constraints that can be derived from the chemical abundance observations. In this work, we investigate the high values of the S/O ratio and its trend with O/H that have been observed at very low metallicity. We have checked if the observed S/O ratios, which do not follow the expected constant behaviour considering a similar nucleosynthetic origin for both S and O, could be explained from the point of view of stellar evolution. 
The peculiarly high S/O ratios seen at low metallicity ($\lessapprox$ 0.1 Z$_{\odot}$) show a different trend compared to galaxies at high 12+log(O/H),  
as presented in Figure~\ref{soratio}. To test this behaviour we start by applying a base model as M1 which was previously used by \cite{Goswami2020} to explain the data in the high metallicity range following the chemical evolution pattern of the  Milky Way (MW) disc and other local galaxies. Model M1 has a canonical \cite{1993MNRAS.262..545K} IMF with an upper mass limit of 100 M$_{\rm \odot}$,  and includes all the parameters required to reproduce the chemical evolution of the MW \citep{Grisoni_2017,Grisoni2018,Grisoni2019,Grisoni2020a,Spitoni2021,Goswami2020}. As illustrated in Figure~\ref{soratio}, model M1, shown in grey dashed line, goes through the HII Gal sample and the majority of the Izotov sample with higher O/H, but is unable to achieve the high S/O ratios shown by the DHR sample at low metallicities. Since the upper mass limit of the IMF in this model is 100 \Msun, and as explained in Sect.\ref{yields}, till this mass range, CCSN at low metallicities doesn't produce enough S as shown in Fig.\ref{pisnyields}, to increase the S/O ratios. Most of the points at high O/H resemble the abundances and ratios derived in local group galaxies, and our model appears to reproduce this part of the observations. We note that although the HII Gal sample has galaxies with very young ages, this model which doesn't include a starburst phase reaches 12 + log (O/H) $\sim$ 7.5 at 200 Myr shown by the grey square.   This model, if used with an appropriate SFH (e.g. increasing the star formation efficiency) more suitable for starburst-dominated galaxies like in the HII Gal and \cite{Izotov2006} samples, can reproduce the cyan and red dots in Fig.~ \ref{pisnyields} with the appropriate young ages expected for such galaxies. Nonetheless, this is not the goal of this work, and in any case, this reference model shows that with a canonical IMF and without the PISN contribution, the high S/O ratios observed in the DHR sample cannot be obtained.     \\

A similar problem was recently studied by \cite{Goswami2022} where they reproduced peculiarly high Fe/O ratios from EMPGs. These EMPGs follow a decreasing trend of Fe/O with increasing metallicity; this trend has also been observed in the case of the DHR subsample's S/O ratio as shown in Figure~\ref{soratio}. These authors explain how the observed high Fe/O ratios and their decreasing trend with metallicity can be reproduced only by using a top-heavy IMF and taking into account the PISN stage in the chemical evolution modelling. On the other hand, we comment in Sect.\ref{yields} while comparing the shapes of PISN ejecta that S and Fe show very similar trends in their ejecta, in a similar range of masses of stars going through the PISN stage. Thus, we tested three models M3, M5 and M6 (M5 and M6 each with two different rotational parameters $\omega$=$\Omega/\Omega_{crit}$=0.3 and 0.4) used by \cite{Goswami2022} to reproduce the high Fe/O ratios and investigate if these models can explain the high S/O as well. We show in Table \ref{parameter} the characteristics of each model tested in this work and present them here in brief; however, the reader is directed to \cite{Goswami2022} for a detailed description of these models. Model M3 uses a top-heavy IMF with a \MUP=300 \Msun\ and a slope of \xUP=0.6. This model takes into account the PISN stage due to the high \MUP\ which causes the S/O ratio to grow due to increased production of S in the low metallicity region as shown by the yellow dashed-dotted line in Fig.~\ref{soratio}. However, this model, although increasing S/O ratio compared to the HII Gal sample
and going through the blue sample of \cite{Yates2020}, it cannot reproduce the highest S/O values observed (around 12+log(O/H)$\sim$ 7.1
; brown dots) which is the main purpose of this work. The age obtained by the model at 12 + log (O/H) $\sim$ 7.5 is 30 Myr which is much younger compared to M1 at the same metallicity. 
Although both the HII Gal and DHR samples include metal-poor and bursty star-forming regions, which in principle would be favourable locations for very massive star formation, there could be other reasons for the difference in S/O ratios operating at the low metallicity end of the sample. The exact characteristics associated with the formation process of these extremely massive stars are still unknown, as are the environmental circumstances that led to their formation. One possible reason that may explain the difference seen between the S/O ratios of HII Gal and DHR low-metallicity samples, supposedly likely for very massive star formation, could be related to the variations in the maximum mass limit of stars that form in different galaxies.

For instance, a star in the low-mass PISN regime in Figure 1 with a mass of 150 \Msun\ produces more O than S as it doesn't reach high enough temperatures to fuse heavier elements, resulting in a lower S/O ratio. On the other hand, a star in the high-mass PISN regime, mass 250 \Msun\, produces more S than it did previously, meaning that the S/O ratio would be higher.
Moreover, \cite{Chru2023}  has demonstrated that distinct trends in the element ratios may be produced associated with specific SFR (sSFR). They discovered a distinct [O/Fe] ratio evolution with the sSFR; however, the PISN did not account for this. Such results would point towards a further enhancement of this dependency. Additionally, further study of these environments
could yield significant information about characteristics other than
low-metallicity and bursty star formation regions, which could be
favourable to the creation of VMS. This will allow for more constraints on the formation of VMS and other nucleosynthesis channels. Nonetheless, with the current constraints available on the abundance ratios, the ages and star formation characteristics,  the PISN channel could not be ruled out.

In models M5 and M6 a bi-modal IMF has been assumed to sample the effects of VMOs exploding as PISN. To do so,  in the chemical evolution model, an early phase with a top-heavy IMF, suitable for Pop III-like stars \citep{wiseetal03}, has been used which has the following form:
\begin{equation}\label{eqwise}
\phi(\Mi) =  \frac{dn}{d\log(\Mi)} \propto \Mi^{-1.3}\times~exp\left[\,- \left(\, \frac{\Mchar}{\Mi} \right)^{1.6}\right]\,  
\end{equation}
A \Mchar\ of 200 M$_{\rm \odot}$ has been adopted for this IMF instead of \Mchar\ of 40 \Msun used in the original work since the ejecta of S and O increase strongly with the initial masses in the high mass range as shown in Sect. \ref{pisnyields}. We also checked with \Mchar\ of 300 M$_{\rm \odot}$ and since the results were similar to \Mchar\ = 200 M$_{\rm \odot}$, we only show here one case. The PISN enrichment phase is limited to the early evolution of the starburst, the duration of which is constrained by the critical metallicity, \Zcr.  When this gas metallicity is reached, a Kroupa IMF (equat. \eqref{imfkroupa}) with \xUP=1.3 and \MUP=40 M$_{\rm \odot}$ is adopted. Since higher mass loss rates cause stars to be unable to reach high enough masses to enter the PISN regime as metallicity grows, an IMF change with a lower \MUP\ is adopted. The critical metallicity is \Zcr\ $=$ 10$^{-2}$ for M5 ( \Mcot\ and \Mcoq) and \Zcr\ $=$ 10$^{-4}$ for M6 (\Msot\ and \Msoq) depending on whether dust cooling is efficient or not \citep{Schneider2006MNRAS,salvadorieial2008MNRAS}. To model the star formation, a star formation efficiency of $\nu$=1 Gyr$^{-1}$ has been adopted for M5. In the left panel of Figure~\ref{soratio} model M5 is shown in black dotted (\Mcot) and dashed lines (\Mcoq) for two different values of rotation considered for the PISN yields. This model reaches the highest S/O ratios at around 15 Myr and to be consistent with the young ages of these galaxies, these models have been shown until 200 Myr of their evolution, unlike models M1 and M3 where their entire evolution (13 Gyr) is shown. 

The model can reach the high S/O ratios at lower 12+(O/H) where an initial burst is modelled that takes into account the presence of VMOs and, once the critical metallicity is reached and the effect of VMO enrichment ceases and the IMF is switched to  \cite{1993MNRAS.262..545K} like resulting in the decreasing S/O ratio present in the observations. 
The model starts with a metal-free gas, Z = 10$^{-10}$ and, as the VMOs start to explode, the metallicity rapidly increases and reaches the \Zcr\ value in a  short period (see Figure 4 of \citealp[]{Goswami2022}). During this phase, the PISN contributes to increased sulfur production, and due to this, the model can reach high S/O ratios. Once the \Zcr\ is reached, the IMF changes to a Kroupa-like IMF where stars are only formed till with \Mi $\leq$ 40 \Msun. This causes the S/O ratio to decrease with increasing metallicity, but the model does not go through the  DHR sample points shown in brown at higher metallicities and results in slightly higher ages  as expected from regions dominated by very massive stars, as shown by \cite{Schneider2006MNRAS}. 

\begin{table}
\small
\caption{Chemical evolution and IMF parameters of the selected models used in this work. These models are shown in Fig. \ref{soratio}.}
\label{parameter}
\centering
\tiny
\begin{tabular}{|c|c|c|c|c|c|c|c|}
\hline
\multicolumn{1}{|c|}{ Model } &
\multicolumn{5}{|c|}{Chemical evolution parameters} &
\multicolumn{2}{|c|}{ IMF }  \\
\hline
 & $\nu$ & $k$ & $\tau_{\rm inf}$ & $A_{\rm SNIa}$ & $\Zcr$ & $\MUP$ & $\xUP$  \\
 & (Gyr$^{-1}$) & & (Gyr) & & & & \\
\hline
M1 & 0.8 & 1.0 & 6.0 & 0.04 & & 100 & 1.5  \\
\hline
M3 & 0.3 & 1.0 & 1.0 & 0.04 & & 300 & 0.6 \\ 
\hline
M5  & 1.0 & 1.0 & 0.1 & 0.04 & 0.01 & 200 &  bi-modal \\
\hline
M6  & 5.0 & 1.0 & 0.1 & 0.04 & 0.0001 & 200 & bi-modal \\
\hline
\end{tabular}
\end{table}

In order to explain this decreasing trend of the DHR sample  as the metallicity increases while being consistent with younger ages, we use Model M6, which is modelled with two different values of rotational parameters ($\omega$=0.3,0.4) and a more efficient star formation and is shown in the right panel of Figure~\ref{soratio} in the black dotted (\Msot) and black solid lines (\Msoq).  This model can not only match the highest S/O ratio shown by the DHR sample at low O/H  but it also shows the decline of the S/O ratio as we go to higher metallicities, although with a steeper slope.  To reproduce this decline of S/O we adopt this model, where once it reaches the \Zcr\, it takes into account the stars that overproduce oxygen or underproduce sulfur, as shown by the S/O ratio of the DHR data in the high metallicity regime. Once the critical metallicity is reached, a Kroupa IMF with \MUP = 150 M$_{\rm \odot}$ is used.  In this mass range, stars produce large amounts of oxygen as compared to sulfur, which causes the S/O ratio to decrease. 
Therefore, we have shown that, by adopting an initial burst of VMOs and using a bi-modal IMF both, the high S/O ratios of the DHR subsample and then the decline with increasing metallicity observed can be explained.   Since this model uses a higher star formation efficiency of $\nu$ =5 Gyr$^{-1}$ combined with a top-heavy IMF, the ages obtained at the same metallicity compared to model M5 are lower. The two black squares correspond to ages 5 and 15 Myr and this model reaches the highest S/O ratios before 5 Myr which is consistent with ages derived from regions in the Magellanic Clouds hosting very massive stars \citep{Schneider2006MNRAS,Crowtheretal2016MNRAS}.
Hence we demonstrate that including PISN in the models could be a possible scenario to explain these high S/O ratios. \cite{Diaz2022} suggested in their comprehensive and inspiring work that the decline in the S/O ratio with increasing metallicity might be related to the chemical evolution of massive star nucleosynthesis 
which produces more sulfur entwined with variations in the IMF which we have investigated through this model.They also comment that sulfur production from SNe Ia is insufficient to produce the desired slope for the S/O decline \citep{Weinberg2019}. We note that to model the exact slope of the decline using the two datasets shown in the DHR sample of \cite{Diaz2022} and \cite{Yates2020} will require a homogeneous derivation of S/O abundances between the samples and obtaining more data, especially for the low metallicity regions, which would be the aim of a future study. This would allow us to get a constraint on the shape of the bi-modal IMF once \Zcr\ is reached.  

\section{Conclusions and outlook}
\label{conclu}
In this paper, we use chemical evolution models to analyze the observed sulfur and oxygen abundances derived in
large samples of star-forming galaxies 
which indicate higher S/O ratios at low metallicities, deviating from a constant solar S/O ratio suggested by other works. To do so we use chemical evolution models as a tool to investigate under which evolutionary scenario these high S/O ratios can be explained. 

Similar peculiar high Fe/O ratios at low metallicities have been reported for EMPGs (e.g. \citealt{Kojima2020,Izotov2006}). \cite{Goswami2022} reproduced these high Fe/O ratios taking into account the later evolutionary stages of massive stars such as PISN and PPISN in the chemical evolution modelling. We compare the stellar yields of S and Fe in the VMO mass range from the yield set used in their work.  In Figure~\ref{pisnyields} we show that the yields of S and Fe follow a similar trend in this mass range where VMOs go through the PPISN/PISN process. We also show that at low metallicities, the sulfur and oxygen production through the CCSN channel is significantly less than the PISN channel. Additionally in Fig. ~\ref{imfyields} we show the IMF-averaged ratios of S/O and Fe/O obtained for two different IMFs to show the differences between only CCSN production and including PISN channels. We thus have used the models that reproduce the observed high Fe/O ratio of these galaxies to test if, considering these later evolutionary stages as a possible channel, the high S/O ratios shown by the DHR subsample could also be simultaneously explained. Firstly, we show that using model M1 previously applied to reproduce the MW disc with a canonical IMF, the high S/O ratio at low metallicities cannot be reproduced. This model does not take into account the PISN stage and has an upper mass limit of \MUP=100 \Msun\  does not produce enough sulfur  through the CCSN channel at low metallicities to obtain a high S/O ratio. It can, however, explain the S/O ratios shown by the HII Gal and Izotov samples derived at higher 12+(O/H) albeit with a higher age, since in this metallicity range the data appear to follow the trend of the local galaxies. 

Thereafter, we used models that take into account stars going through the PISN stage. Model M3 uses a top-heavy IMF with a higher \MUP\ of 300 \Msun\ and slope of \xUP=0.6, increasing the S/O ratios at lower metallicities to go through the \cite{Yates2020} sample but not enough to reach the highest S/O values observed in the DHR sample of \cite{Diaz2022}. Subsequently, to properly account for the PISN evolutionary stage and reach the high S/O ratios, we use a bi-modal IMF where the first part takes into account an IMF devised for Pop III-like stars \citep{wiseetal03} with a burst that includes the effects of VMOs and we limit this early phase using a critical metallicity, \Zcr\ since VMOs only form in low metallicity environments according to theoretical predictions. Once this \Zcr\ is reached, a canonical \cite{1993MNRAS.262..545K} IMF is adopted. The values of \Zcr\ used here are \Zcr =0.0001 and 0.01 depending on inefficient or efficient dust cooling \citep{Schneider2006MNRAS,salvadorieial2008MNRAS}.  In Figure~\ref{soratio}, we show that model M5 at low metallicity can reach the high S/O values as indicated in the DHR sample. In this model, the early phase takes into account a strong burst for which a star formation efficiency of $\nu$ =1 Gyr$^{-1}$ has been used to reach a high S/O ratio at young ages. During the PISN stage, sulfur production increases as compared to oxygen in the high-mass PISN regime. Hence a high S/O ratio can be obtained. Once the critical metallicity is reached, VMOs stop being formed, leading to a reduction in sulfur production, and we can see the decreasing trend of S/O shown by the data being reproduced by the model as well, although with a different slope. The effect of rotation on the production of sulfur is minimal, as can be seen in Figure~\ref{soratio} where the two models with different rotational ($\Omega$) values follow roughly the same path. 

Since model M5 albeit going through the highest S/O ratios at low metallicity,  does not go through the DHR sample as the metallicity increases, combined with slightly higher ages as expected from regions hosting very massive stars, we used model M6 which has a higher star formation efficiency of $\nu$ =5 Gyr$^{-1}$ to be able to reach the high S/O ratio at a very young age consistent with previous studies. Model M6 like M5 can reach a high S/O ratio at low metallicity because of the increased production of sulfur from the PISN stage; then the decreasing trend of the S/O starts once the model reaches \Zcr .  In this case, because of the higher efficiency in star formation and populating with more stars that produce more oxygen than sulfur using an appropriate IMF, for which a second \cite{1993MNRAS.262..545K} IMF with an \MUP=150 M$_{\rm \odot}$ and \xUP =0.9 has been adopted, the decrease in S/O observed is also shown by the model with a stepper slope than observed.  In order to accurately replicate the slope, more low metallicity data demonstrating these high S/O ratios would need to be obtained, along with a uniform derivation of the S/O ratios of the two samples of DHR and \cite{Yates2020}. This would be the aim of a subsequent study, and in this instance, we offer the theoretical scenario of this decline.

Hence, we show here that by taking into account later evolutionary stages like the ones of the PPISN/PISN process, and adopting an IMF somewhat more suitable for star formation events as those recently studied in the early universe, the observed high S/O ratios at low metallicity can be reproduced.  Such top-heavy IMF with \MUP\ of 200 M$_{\rm \odot}$ as reported for 30 Doradus and, in general, in other bursty star-forming regions, has been predicted which favours the formation of such very massive stars \citep{marks_imf_2012MNRAS,jerabkova2018,Zhang2018Natur,Crowther2010,Schneider2018}. Hence PISN provide a possible channel to obtain high S/O ratios in low-metallicity star-forming objects.

However, due to the lack of accurate derivations of sulfur abundances in the low metallicity range, the IMF slope and the upper mass limit are difficult to constrain robustly. This will also allow us to find models that fit these datasets better which is the purpose of a future study, as discussed before.   
To elucidate the characteristics of this intriguing scenario of chemical enrichment from very massive stars, additional observations and analysis of other elemental abundance ratios are required. Moreover, this would enable us to distinguish between various possible nucleosynthesis channels.  With more data coming from JWST to derive chemical abundances, including sulfur, of young star-forming galaxies we could try to constrain the early chemical enrichment scenario in a more robust way. \\

 In this work, we have shown that with the study of the chemical abundance and the S/O ratio of star-forming galaxies, important properties such as the mass limit of the very massive stars can be better theoretically constrained, which could help in e.g. compelling the upper mass limit and tail of the IMF in the early universe. Similarly, the high Fe/O ratios that were observed in these galaxies at low metallicity have been recently explained by theoretical models that include the PISN stage \citep{Goswami2022}. However, iron lines are typically faint or undetectable in this class of objects, whereas sulfur lines are frequently seen in their spectra even in high redshift galaxies. In addition, sulfur, unlike iron, does not suffer high depletion in dust grains. 
To this end, we may encourage JWST observations of IR emission lines of sulfur ([SIII], [SIV]) and neon ([NeII], [NeIII]), seen in the spectra of star-forming objects, in order to consistently derive S/Ne as a possible proxy for S/O.
 We propose also that the derivation of sulfur abundances and S/O ratios of more distant star-forming galaxies (e.g. with JWST given its spectroscopy performance and timing of data gathering), could pave a new, unexplored path to understanding the role of very massive stars in the early stages of galaxy formation. 

\begin{acknowledgements}
S.G. acknowledges financial support from the European Union under the 2014-2020 ERDF Operational Programme and by the Department of Economic Transformation, Industry, Knowledge,  and Universities of the Regional Government of Andalusia through the FEDER-UCA18-107404 grant. S.G. acknowledges funding from Project CRISP PTDC/FIS-AST-31546/2017 funded by FCT. JVM, BPD, EPM acknowledge financial support from grants CEX2021-001131-S and PID2019-107408GB-C44 funded by MCIN/AEI/ 10.13039/501100011033.

\end{acknowledgements}

%
%

\bibliographystyle{aa}
\bibliography{biblio}

\end{document}